\documentclass[aps,prc,twocolumn,showpacs,nofootinbib]{revtex4-1}
\usepackage[utf8]{inputenc}
\usepackage{graphicx}   
\usepackage{latexsym}   
\usepackage{enumerate}
\usepackage{rotating,booktabs,multirow}
\usepackage{amsmath}
\usepackage{amsfonts}
\usepackage{amssymb}
\usepackage{multirow}
\usepackage{bm}
\usepackage{xcolor}

\begin{document}
\title{Probing system size dependence at high baryon density by systematic comparison of Ag+Ag and Au+Au reactions at 1.23$A$ GeV}
\author{Tom~Reichert$^{1,3}$, Apiwit~Kittiratpattana$^{1,4}$, Pengcheng~Li$^{1,5,6}$, Jan Steinheimer$^{7}$, Marcus~Bleicher$^{1,2,3}$}

\affiliation{$^1$ Institut f\"ur Theoretische Physik, Goethe Universit\"at Frankfurt, Max-von-Laue-Strasse 1, D-60438 Frankfurt am Main, Germany}
\affiliation{$^2$ GSI Helmholtzzentrum f\"ur Schwerionenforschung GmbH, Planckstr. 1, 64291 Darmstadt, Germany}
\affiliation{$^3$ Helmholtz Research Academy Hesse for FAIR (HFHF), GSI Helmholtz Center for Heavy Ion Physics, Campus Frankfurt, Max-von-Laue-Str. 12, 60438 Frankfurt, Germany}
\affiliation{$^4$ Suranaree University of Technology, University Avenue 111, Nakhon Ratchasima 30000, Thailand}
\affiliation{$^5$ School of Nuclear Science and Technology, Lanzhou University, Lanzhou 730000, China}
\affiliation{$^6$ School of Science, Huzhou University, Huzhou 313000, China}
\affiliation{$^7$ FIAS, Ruth-Moufang-Str. 1, D-60438 Frankfurt am Main, Germany}

\begin{abstract}

We present UrQMD predictions for the comparison of the recently measured Ag+Ag and Au+Au runs at a beam energy of $E_\mathrm{lab}$ = 1.23~$A$GeV explored by the HADES experiment. To this aim, different centrality definitions are investigated: We suggest that in the case of particle production, both systems should be compared at the same number of participating nucleons, while for a comparison of the (elliptic) flow, a selection on the initial state eccentricity - as in high energy reaction - is better suited. Generally, we find good agreement between both system, if these centrality criteria are used. Specifically, the deuteron yields per participant and the pion to proton ratios are shown to scale with $A_\mathrm{part}$ except for very central Ag+Ag reactions due to stronger stopping in such reactions. The elliptic flow in both systems follows inital state eccentricity scaling, albeit with the opposite sign as compared to high energies, suggesting a strong relation between final flows and the initial state also at the low energies explored here. The observation of this scaling might then allow to obtain further information on the expansion properties (and the EoS) of matter at high baryon densities. 
\end{abstract}

\maketitle

\section{Introduction}
With the advent of gravitational wave astronomy, the interest into the precise determination of the equation-of-state of Quantum Chromo-Dynamics (QCD) at high baryon densities has advanced to a new level. For the first time, the use of gravitational waves allows for an unprecedented accuracy of the extraction of the properties of neutron stars, and especially their equation-of-state from astrophysical observations \cite{Raaijmakers:2019dks}. This high quality cosmic data calls for further studies of the properties of nuclear matter in earth-bound accelerator experiments. One of the key facilities to explore QCD-matter at similar densities and temperatures is the GSI-SIS18 facility in Darmstadt \cite{Durante:2019hzd}. There, light and heavy atomic nuclei are collided with beam energies from a few hundred $A$MeV up to 2 $A$GeV, recreating the baryon densities ($\rho_B/\rho_0\approx 2-5$) found in the interior of neutron stars. The HADES experiment at GSI has recently published data for Au+Au reactions at $E_\mathrm{lab}$ = 1.23 $A$GeV \cite{HADES:2020lob,HADES:2022osk}. This data indicated the creation of very high baryon densities which needed a hard EoS to allow for a description of the flow data up to the $4^{th}$ flow harmonic \cite{Hillmann:2018nmd,Hillmann:2019wlt,Mohs:2020awg}. On the experimental side, even higher flow harmonic up to $5^{th}$ and potentially $6^{th}$ order could be analyzed, but still wait for theoretical investigation. While a wealth of data has been available from the HADES and previously FOPI experiments, most of these data where unfortunately taken at different beam energies. Another complication emerged from the tension between the FOPI and the HADES data in terms of pion and cluster multiplicities. This called for a comprehensive analysis of the data taken in one single experiment. Just now, the comparison to a lighter system at the same collision energy has  become possible with the new experimental runs of Ag+Ag at $E_\mathrm{lab}$ = 1.23 $A$GeV taken by the HADES experiment. This opens the possibility for a direct comparison of the EoS in the baryon rich regime between a light and heavy collision system.

\section{Simulation set-up}
For this study we use the Ultra-relativistic Quantum Molecular Dynamics (UrQMD) model \cite{Bass:1998ca,Bleicher:1999xi,Bleicher:2022kcu} in its most recent version (v3.5). UrQMD is a dynamical microscopic transport simulation based on the explicit propagation of hadrons in phase-space. In its current version, UrQMD includes a broad variety of baryonic and mesonic resonances up to masses of 4~GeV. First predictions for bulk observables in Ag+Ag at 1.58~AGeV kinetic beam energy have been recently published in Ref. \cite{Reichert:2021ljd} comparing a wide range of transport approaches available for this energy. For the present study, we also include deuterons which are calculated via a coalescence approach in the final state \cite{Sombun:2018yqh}. A recent study has shown that this approach is consistent with MST calculations for cluster production \cite{Kireyeu:2022qmv}. UrQMD has already been successfully applied to calculate direct and elliptic flow and higher order flow components \cite{Hillmann:2018nmd,Hillmann:2019wlt,Reichert:2022gqe} of nucleons and light clusters in the investigated energy regime and has recently demonstrated its abilities to further describe larger clusters ($t$, $^3$He) \cite{Hillmann:2021zgj}.

In this study a nuclear matter equation-of-state following a Skyrme-type potential \cite{Skyrme:1959zz} of the form shown in Eq. \eqref{eq:skyrme} is employed: 
\begin{equation}\label{eq:skyrme}
    U_{\rm Skyrme} = \alpha \left(\frac{\rho}{\rho_0}\right) + \beta \left(\frac{\rho}{\rho_0}\right)^\gamma
\end{equation}
Based on the overall good agreement between our previous studies \cite{Hillmann:2018nmd,Hillmann:2019wlt} in comparison to experimental data \cite{HADES:2020lob} we choose the same hard EoS with the corresponding parameter set shown in Tab. \ref{tab:parameters}. We will not show results for the soft EoS, because the soft EoS did not allow for a good description of the flow data.

UrQMD does further allow to use also more complex equations-of-state, e.g. recently a Chiral Mean Field (CMF) based EoS was used in UrQMD to explore a more complicated density dependence in comparison to the Skyrme-EoS and the onset of deconfinement \cite{Kuttan:2022zno}.

\begin{table} [h!]
    \centering
    \begin{tabular}{c||c}
        Parameter & Hard EoS \\
        \hline \hline
        $\alpha$ & $-124$ MeV \\
        $\beta$ & $71$ MeV \\
        $\gamma$ & $2$  \\
    \end{tabular}
    \caption{Parameters for the hard Skyrme EoS employed in UrQMD.}
    \label{tab:parameters}
\end{table}

\section{Event class selection}
A priori it is not clear how to compare different collision systems in a meaningful way, even at the same collision energy. A wide variety of different methods are possible, e.g. one can compare different systems at the same impact parameter, same participant nucleon number, same charged particle density, same energy deposition, same number of binary collisions, same overlap shape, etc. \cite{WA98:2000mvt,Eyyubova:2021ngi}. For the present paper, we will employ two different methods, namely a selection on the number of participating nucleons or a selection on the same initial transverse overlap eccentricity.

Let us shortly illustrate the reasoning behind these choices: At the rather low energies discussed here, pions are the most abundantly produced secondary particles. Pions stem mainly from the decay of Baryon resonances and therefore one expects a linear scaling of pion production with the number of participating nucleons \cite{FOPI:1997qpm,HADES:2020ver}. 

In case of the flow observables, the number of participants is essentially irrelevant, because the elliptic flow is driven by the spatial anisotropy (and the resulting pressure gradients) which is related to the spatial anisotropy. At high energies there is a strong correspondence between the initial state spatial anisotropy and final state anisotropic flow in momentum space \cite{Bhalerao:2005mm}. At the low energies discussed here, it will be important to explore, if such a correspondence is also present or if spectator shadowing will lead to a substantial breaking of the scaling.

\begin{figure} [t!]
    \centering
    \includegraphics[width=\columnwidth]{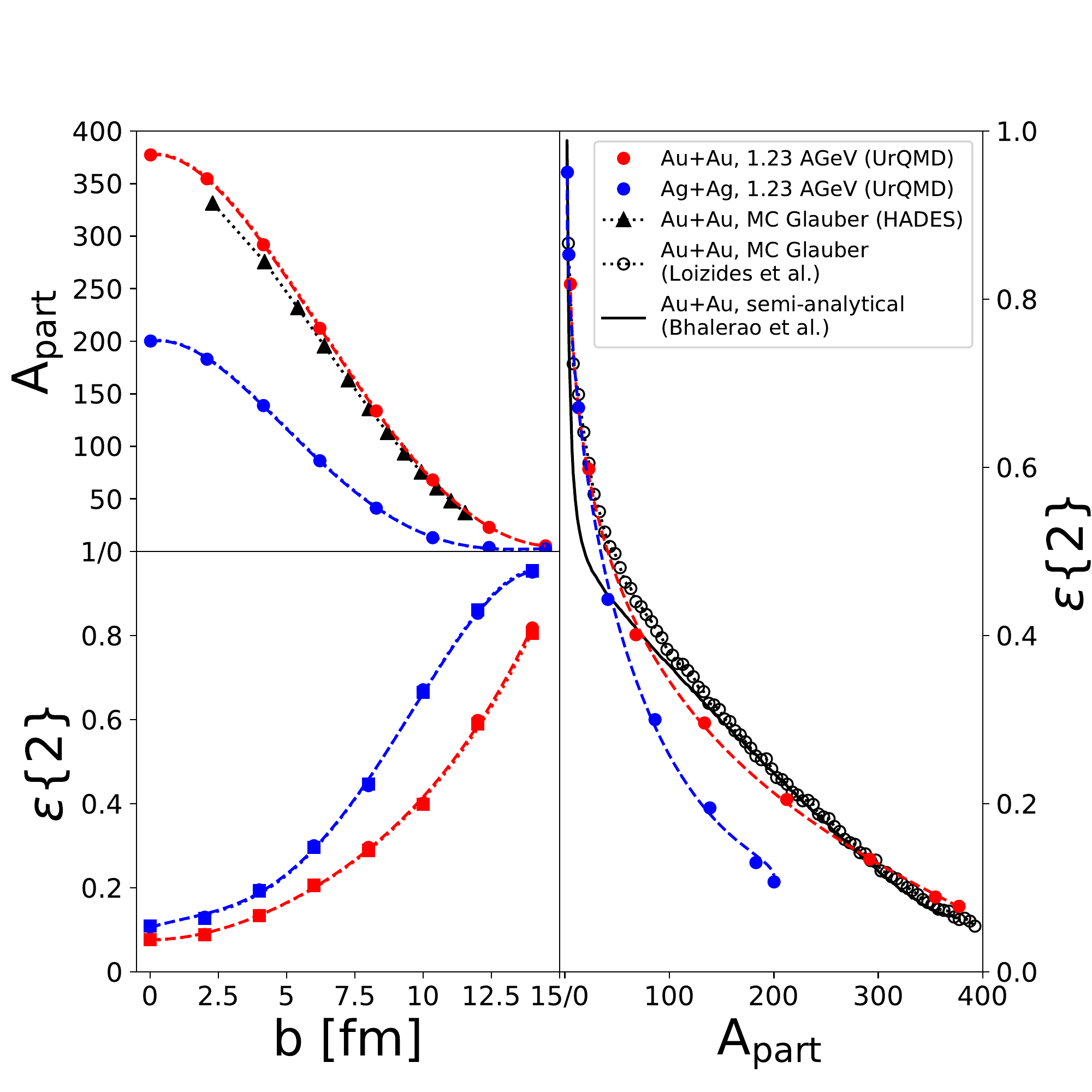}
    \caption{[Color online] Relation between the impact parameter $b$, the average number of UrQMD Monte Carlo Glauber participants $\langle A_{\rm part}\rangle$ (top left) and the average transverse overlap eccentricity $\langle\varepsilon\{2\}\rangle$ (bottom left) for Au+Au collisions (red) and Ag+Ag collisions (blue) from UrQMD in comparison to the MC Glauber results from HADES \cite{HADES:2017def} (black triangles). On the right side, a comparison of the obtained UrQMD Monte Carlo Glauber eccentricities to the results from Ref. \cite{Loizides:2014vua} (black circles) and Ref. \cite{Bhalerao:2006tp} (black line) are shown in comparison to the blue and red circles from the UrQMD simulation.}
    \label{fig:b_Apart_ecc2}
\end{figure}

To make an event/centrality selection that is comparable to that of the HADES experiment \cite{HADES:2017def} who used a Monte Carlo Glauber model \cite{Broniowski:2007nz}, we employ the UrQMD model also in Monte Carlo Glauber mode by only allowing elastic zero-degree scatterings between the impinging nucleons\footnote{The selection of specific centralities and their relation to the number of participants is not straightforward. E.g. at a fixed impact parameter one may use different methods to estimate the number of participants which yield different results. One can define the number of ``Glauber participants" (this is what we use here) which is given be the nucleons in the overlap region, this number is different from the participants one obtains from the simulation, if one counts all nucleons that had interacted, another different number of participants may be obtained if one counts the spectators (nucleons being under small angles) and calculates $A_{\rm part}= 2A-2 A_{\rm spec}$. While all definitions coincide for very central and very peripheral collisions, for mid-central reactions the difference can be comparatively large.}. We will use the mean number of participants for a given impact parameter as estimated from the UrQMD Monte Carlo Glauber calculation to relate the impact parameter $b$ to $A_{\rm part}$. Here (and in the rest of the manuscript), the number of participants is therefore the Monte Carlo Glauber value for $A_\mathrm{part}$ as obtained from the UrQMD in Glauber mode which might differ from the number of participants in the full microscopic simulation which usually is larger due to angular dependent interactions. For the comparison of the flows of the two systems the same UrQMD Monte Carlo Glauber mapping is employed to relate $b$ to the initial transverse overlap eccentricity\footnote{We use the definition of the eccentricity following $\varepsilon\equiv\varepsilon\{2\}=\sqrt{\langle\varepsilon_{\rm part}^2\rangle}$ with $\varepsilon_{\rm part}=\frac{\sqrt{(\sigma_y^2-\sigma_x^2)^2+4\sigma_{xy}^2}}{\sigma_y^2+\sigma_x^2}$ and $\sigma_i^2$ being the (co-)variances.} $\varepsilon$. Here, the eccentricity is calculated from the initial spatial (transverse) distribution of those nucleons which participate in the UrQMD Monte Carlo Glauber calculation. Due to the rather low collision energy the nuclei are not as strongly Lorentz contracted as e.g. at RHIC and therefore we call $\varepsilon$ the ``transverse overlap" eccentricity. The relations between $b$, $\langle A_{\rm part}\rangle$ and $\langle\varepsilon\{2\}\rangle$ at $E_\mathrm{lab}$ = 1.23~$A$GeV are given in Tab. \ref{tab:b-a-eps}. For the specific relations between the impact parameters, eccentricities and the number of participant nucleons we use the fitting function:
\begin{equation}
    \left.{\langle\varepsilon\{2\}\rangle(b) \atop \langle A_{\rm part}\rangle(b)}\right\}= a_4 b^4+a_3 b^3+a_2 b^2+a_1 b^1+a_0
\end{equation}
with the impact parameter $b$ given in fm. For the eccentricities the parameters are: 
$a_4=1.45 \cdot  10^{-5}, a_3=-2.50  \cdot 10^{-4}, a_2=4.49 \cdot  10^{-3}, a_1=-4.57  \cdot 10^{-4}, a_0=7.67  \cdot 10^{-2}$ for Au+Au and 
$a_4=-7.60  \cdot 10^{-5}, a_3=1.66 \cdot  10^{-3}, a_2=-5.64  \cdot 10^{-3}, a_1=2.17 \cdot  10^{-2}, a_0=1.05  \cdot 10^{-1}$ for Ag+Ag. 
For the participants the parameters are: 
$a_4=-1.15  \cdot 10^{-2}, a_3=5.93 \cdot  10^{-1}, a_2=-8.11, a_1=2.34, a_0=377.42$ for Au+Au and 
$a_4=-1.88  \cdot 10^{-2}, a_3=6.92 \cdot  10^{-1}, a_2=-7.28, a_1=3.83, a_0=199.95$ for Ag+Ag. The values obtained in this work are consistent with previous estimates of the HADES experiment\footnote{Note, the actual $A_{\rm part}$ values after the full UrQMD simulations do not coincide with the Glauber model $A_\mathrm{part}$ used here. Also due to the long interpenetration time it is actually questionable to define an transverse overlap eccentricity at any given time. What is done here is to make the model simulations comparable to the experimental analysis.} and also coincide with results from Refs. \cite{Alver:2008aq,Loizides:2014vua,Bhalerao:2006tp,HADES:2017def}. Our simulations and the previous results are provided in Fig. \ref{fig:b_Apart_ecc2}, which shows the relation between the impact parameter $b$, the average number of UrQMD Monte Carlo Glauber participants $\langle A_{\rm part}\rangle$ and the average transverse overlap eccentricity $\langle\varepsilon\{2\}\rangle$ for Au+Au collisions (red) and Ag+Ag collisions (blue) from UrQMD in comparison to the Monte Carlo Glauber results from HADES \cite{HADES:2017def} (black triangles). Also shown are results from Ref. \cite{Loizides:2014vua} (black circles) and Ref. \cite{Bhalerao:2006tp} (black line).

\begin{table} [t!b]
    \centering
    \begin{tabular}{l||c|c|c|c}
          \multirow{2}{*}{$b$ [fm]} & \multicolumn{2}{c|}{Au+Au} & \multicolumn{2}{c}{Ag+Ag} \\ \cline{2-5}
          & $\langle A_{\rm part}\rangle$ &  $\langle\varepsilon\{2\}\rangle$  & $\langle A_{\rm part}\rangle$ & $\langle\varepsilon\{2\}\rangle$ \ \\
         \hline \hline
         0 & 377.3 & 0.078 & 200.2 & 0.107 \\ \hline
         2 & 354.6 & 0.089 & 182.9 & 0.130 \\ \hline
         4 & 291.8 & 0.134 & 138.8 & 0.195 \\ \hline
         6 & 212.3 & 0.205 & 86.3 & 0.300 \\ \hline
         8 & 133.7 & 0.296 & 41.2 & 0.443 \\ \hline
         10 & 68.0 & 0.401 & 13.1 & 0.671 \\ \hline
         12 & 22.9 & 0.598 & 3.8 & 0.853 \\ \hline
         14 & 5.3 & 0.818 & 2.4 & 0.951 \\ \hline
    \end{tabular}
    \caption{Relation between impact parameter $b$, the number of participants $\langle A_{\rm part}\rangle$ and the eccentricity $\langle\varepsilon\{2\}\rangle$ for Au+Au and Ag+Ag reactions at $E_\mathrm{lab}$ = 1.23~$A$GeV.}
    \label{tab:b-a-eps}
\end{table}

In order to compare the results from the silver-silver and gold-gold collision systems we then employ the simulations at fixed average number of participants or at fixed average transverse overlap eccentricity. The corresponding impact parameter values for fixed $\langle A_{\rm part}\rangle$ and $\langle\varepsilon\{2\}\rangle$ are shown in Tab. \ref{tab:b-a-eps-fit}. 

\begin{table} [t!h]
    \centering
    \begin{tabular}{c||c|crc||c|c}
          & \multicolumn{2}{c}{b [fm]} & \hspace{1cm} & & \multicolumn{2}{c}{b [fm]} \\
          \cline{0-2}\cline{5-7}
          $\langle A_{\rm part}\rangle$ & Au+Au & Ag+Ag &  & $\langle\varepsilon\{2\}\rangle$ & Au+Au & Ag+Ag \\
          \cline{0-2}\cline{5-7}
          200 & 6.3 & 0.2 &  & 0.2 & 6.0 & 4.4 \\
          \cline{0-2}\cline{5-7}
          150 & 7.6 & 3.6 &  & 0.3 & 8.2 & 6.1 \\
          \cline{0-2}\cline{5-7}
          100 & 9.0 & 5.5 &  & 0.5 & 11.1 & 8.4 \\
          \cline{0-2}\cline{5-7}
           50 & 10.7 & 7.6 &  & 0.7 & 13.1 & 10.3 \\
          \cline{0-2}\cline{5-7}
    \end{tabular}
    \caption{Corresponding impact parameter values for fixed $\langle A_{\rm part}\rangle$ or $\langle\varepsilon\{2\}\rangle$ as obtained from the fit function for Au+Au and Ag+Ag at $E_\mathrm{lab}$ = 1.23~$A$GeV.}
    \label{tab:b-a-eps-fit}
\end{table}

\section{Comparison of rapidity distributions and $p_\perp$ spectra}
\begin{figure} [t!hb]
    \centering
    \includegraphics[width=\columnwidth]{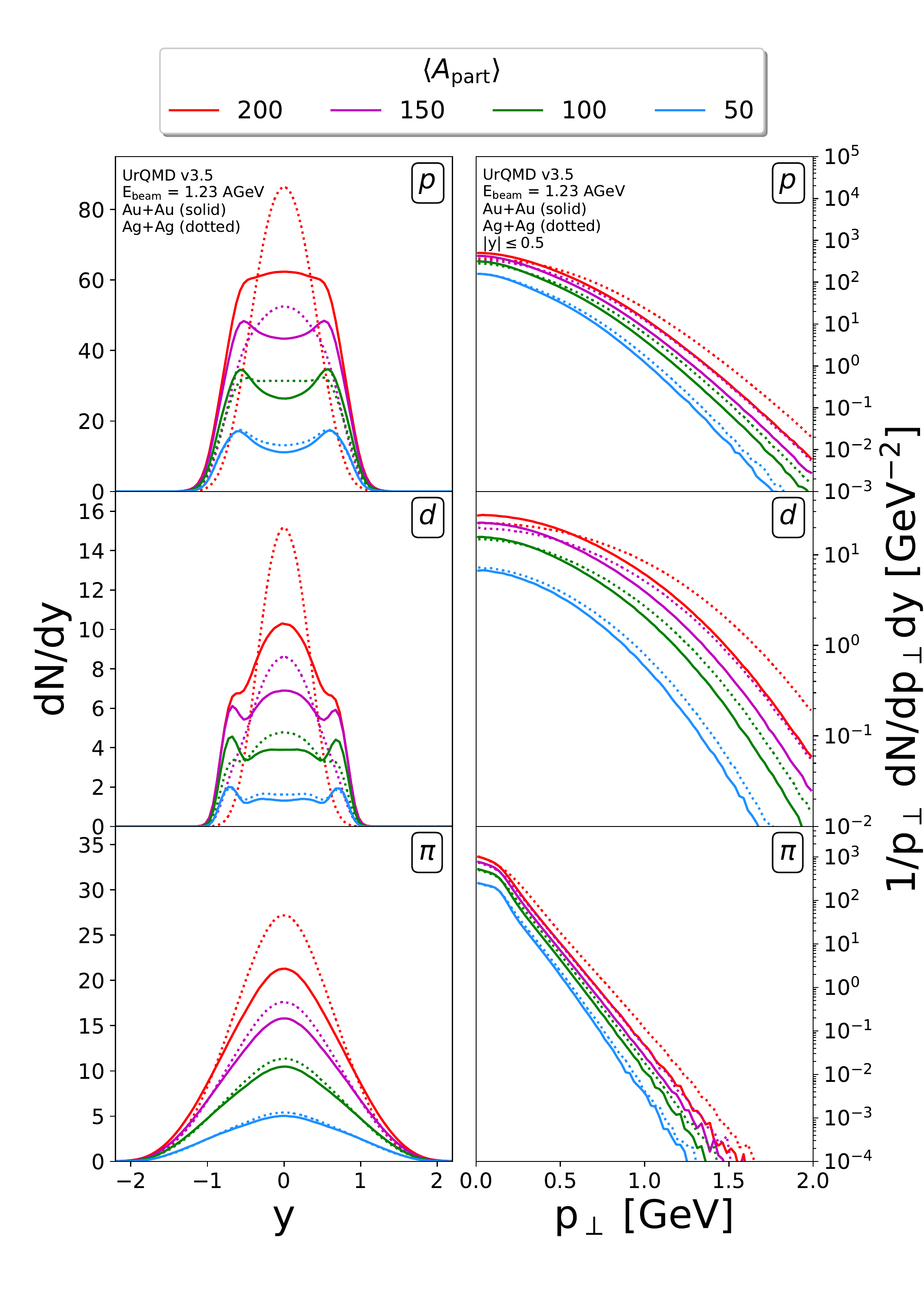}
    \caption{[Color online] The rapidity distributions (left column) and the transverse momentum spectra (right column) of protons (top row), deuterons (middle row) and pions (bottom row) for Ag+Ag reactions (dotted lines) and Au+Au reactions (solid lines) at $E_\mathrm{lab}$ = 1.23 $A$GeV. The average number of participants $\langle A_{\rm part}\rangle$ is fixed to 200 (red), 150 (magenta), 100 (green) and 50 (blue) as defined above.}
    \label{fig:dNdy_dNdpt}
\end{figure}

To set the stage we start with the proton, deuteron and pion distributions in Ag+Ag and Au+Au reactions. We compare Ag+Ag reaction and Au+Au reactions at fixed number of participants for $\langle A_{\rm part}\rangle=200, 150, 100, 50$ as defined above. The left column of Fig. \ref{fig:dNdy_dNdpt} shows the rapidity spectra of participating protons (top left) for Ag+Ag (dotted line) and Au+Au (solid line). It is clear that the fragmentation/spectator regions ($|y|>0.6$) are different, due to the different number of semi-spectators (nucleons that have only undergone a single collision). In addition, we observe that the distributions at midrapidity are very similar for peripheral Ag+Ag and Au+Au reactions. Nevertheless, one clearly observes that very central Ag+Ag reactions are rather different from mid-peripheral Au+Au reactions. The stopping in central Ag+Ag reactions is substantially stronger. We will discuss this phenomenon below in more detail. In the middle left plot we show the deuteron distributions for the same reactions. In line with the differences for the single nucleons one observes also that the deuteron yields at central rapidities in Au+Au are lower than in Ag+Ag (at the same number of participants). Finally, we compare the rapidity spectra of the pions (bottom left plot) in Ag+Ag and Au+Au. Here, we see a slight suppression of the midrapidity pion yields in Au+Au as compared to Ag+Ag. The reason is, again, the weaker stopping in peripheral Au+Au as compared to central Ag+Ag which leads to less pion production at the same number of participants.

\begin{figure} [t!hb]
    \centering
    \includegraphics[width=\columnwidth]{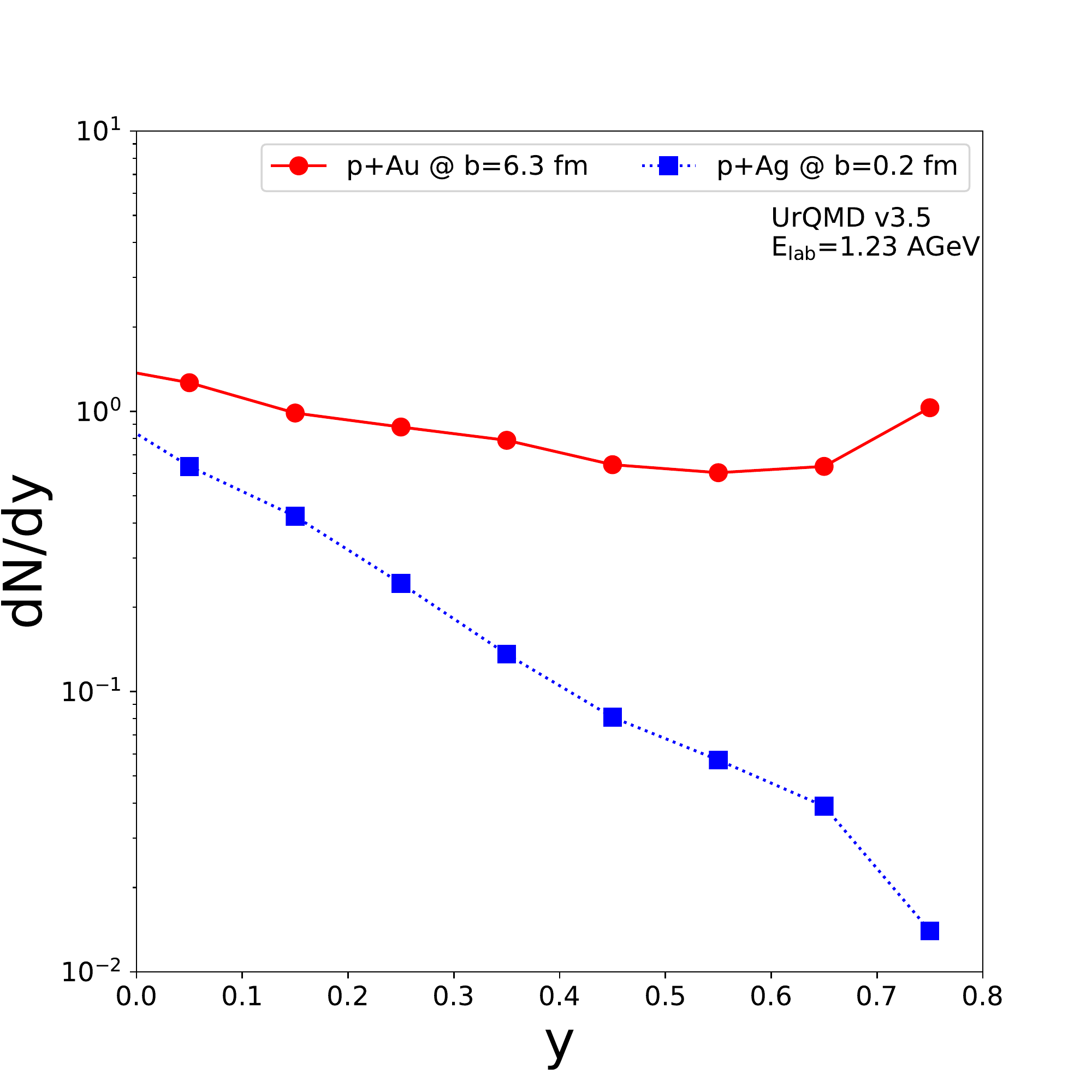}
    \caption{[Color online] Rapidity distributions of the projectile nucleons in p+Au collisions at $b=6.3$ fm (full line, red circles) and in p+Ag collisions at $b=0.2$ fm (dotted line, blue squares) from UrQMD at $E_\mathrm{lab}$ = $1.23A$ GeV kinetic beam energy.}
    \label{fig:stopping}
\end{figure}
Let us now turn to the right side of Fig. \ref{fig:dNdy_dNdpt}. Here the transverse momentum distributions at the same number of participants are compared between Ag+Ag reactions and Au+Au reactions at fixed number of participants for $\langle A_{\rm part}\rangle=200, 150, 100, 50$ as defined above. From top to bottom the right column shows the transverse momentum spectra of protons (top row) for Ag+Ag (dotted line) and Au+Au (solid line) at midrapidity ($|y|<0.5$). We observe a stronger radial flow in the smaller Ag+Ag system as compared to the Au+Au system. The difference is biggest for central Ag+Ag reactions in comparison to semi-peripheral Au+Au reactions (albeit both systems have 200 participants) and vanishes towards more peripheral Ag+Ag and Au+Au reactions at the same number of participants. Again, we relate this difference to the stronger stopping power of the central Ag+Ag reaction in comparison to semi-peripheral Au+Au collisions. In the middle row we show the deuteron distributions for the same reactions. Again, Ag+Ag reactions and Au+Au reactions are rather similar for fixed number of participants, except for the most central case with $A_\mathrm{part}>150$. For the most central Ag+Ag reactions, substantially stronger flow, visible by the shoulder-arm shape of the distribution, is developed than for the Au+Au case. Finally, the comparison of the the transverse momentum spectra of the pions (bottom row) in Ag+Ag and Au+Au is shown. In contrast to the proton and deuteron transverse flow spectra, the shapes are nearly exponential, but again a higher slope is observed for the most central Ag+Ag reaction as compared to the Au+Au collision. 

Why is the stopping and transverse expansion in Ag+Ag and Au+Au so different even for fixed number of participants (especially for $A_\mathrm{part}>150$)? This is elucidated in Fig. \ref{fig:stopping} where we show the rapidity distributions of the projectile nucleons (after the collision) in p+Au collisions at $b=6.3$ fm (full line, red circles) and in p+Ag collisions at $b=0.2$ fm (dotted line, blue squares) from UrQMD at $E_\mathrm{lab}$ = $1.23A$ GeV kinetic beam energy. The impact parameters are the same as for Ag+Ag (Au+Au) at $A_\mathrm{part}=200$. The Figure demonstrates that the rapidity loss of a nucleon \cite{Blume:2008zza} when passing centrally through the Ag-nucleus ($\langle\delta y\rangle = \langle y\rangle - y_\mathrm{initial}=0.55$) is substantially stronger than when passing peripherally through the Au-nucleus ($\langle\delta y\rangle = 0.37$). This stronger stopping power of the Ag-nucleus (when folded with the nucleon distribution of the nucleus in the overlap region) results in more peaked rapidity distribution for Ag+Ag as compared to Au+Au. The system created from a Ag+Ag collision with the same number of participants as an Au+Au collision simply has a more compact geometry. In turn this results in a stronger transverse expansion in the Ag+Ag case which is reflected in the transverse momentum spectrum. In conclusion, one expects a breakdown of scaling when comparing Ag+Ag with Au+Au reactions for $A_\mathrm{part}>150$. 

\begin{figure} [t!hb]
    \centering
    \includegraphics[width=\columnwidth]{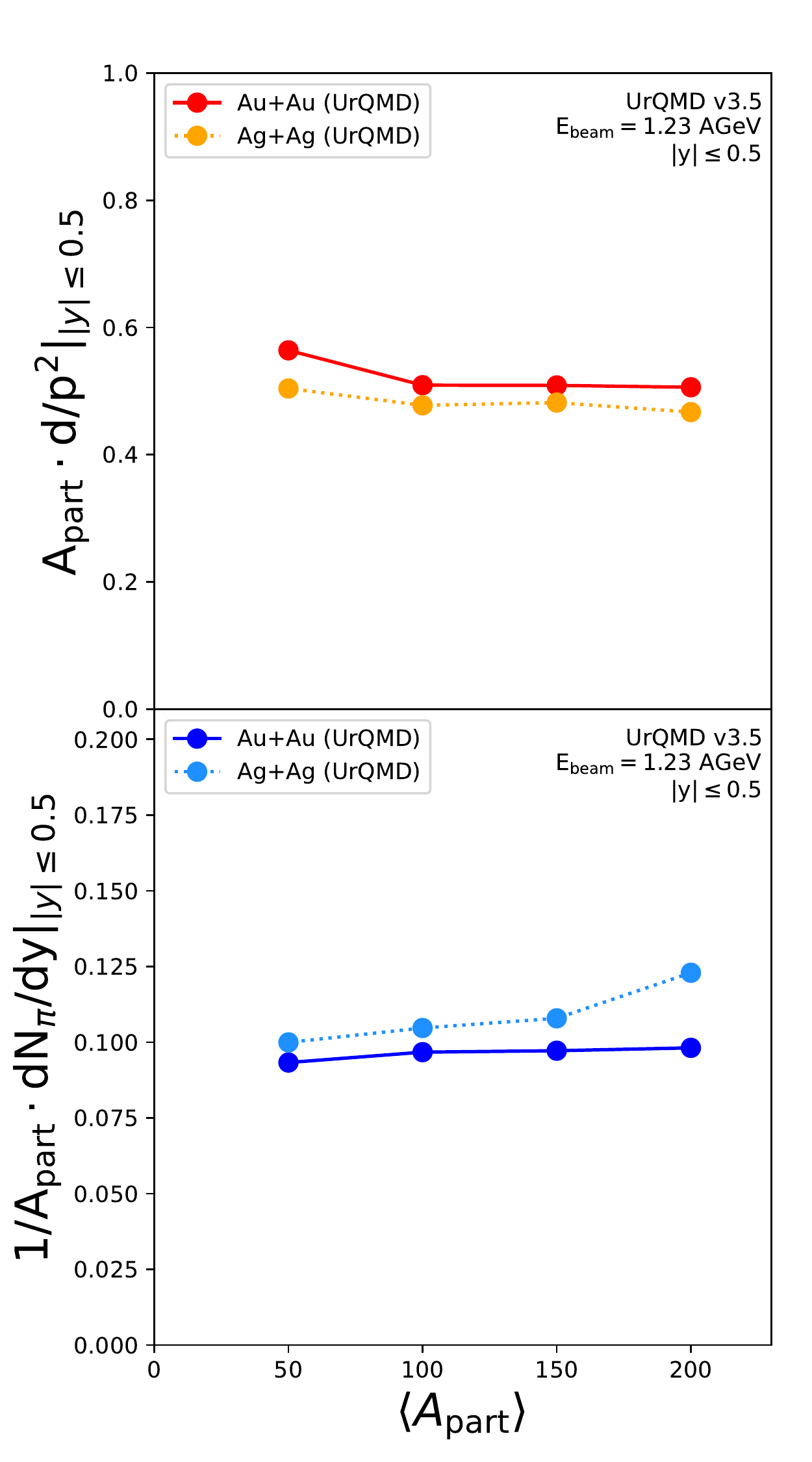}
    \caption{[Color online] The $A_\mathrm{part}\cdot d/p^2$ ratio (top panel) at midrapidity and the number of $\pi/A_\mathrm{part}$ (bottom panel) at midrapidity as a function of $\langle A_{\rm part}\rangle$ for Ag+Ag reactions (dotted lines) and Au+Au reactions (solid lines) at $E_\mathrm{lab}$ = 1.23 $A$GeV from UrQMD.}
    \label{fig:multiplicity_scaling}
\end{figure}

A similar observation is present in the case of the multiplicities of newly produced particles. To quantify the size of this effect, Fig. \ref{fig:multiplicity_scaling} shows in the bottom part the density of pions (sum of all pion states) at midrapidity per participant as a function of the number of participants from UrQMD (Ag+Ag as dotted line, Au+Au as full line). As discussed above, the pion yield scales nearly perfectly with the participant number in both systems for $A_\mathrm{part}<150$, but the Ag-system breaks the scaling for the most central reactions. Having stronger stopping than the gold-gold system, the more compact silver-silver collision at equal number of participants provides more energy for particle production and transverse expansion for $A_\mathrm{part}>150$. The top part of Fig. \ref{fig:multiplicity_scaling} depicts the scaling of deuteron production at midrapidity. Here one can test the well known scaling $d/p^2\propto 1/\mathrm{Volume}$. Under the assumption that the volume is proportional to $A_\mathrm{part}$ one expects a flat line for the ratio $A_\mathrm{part}\cdot  d/p^2$ as a function of the number of participants. Indeed such a scaling is observed for $A_\mathrm{part}>50$ in both collision systems with a minor deviation at very peripheral collisions. 

\section{Comparison of flow harmonics}
Finally, we focus on the discussion of the different flow harmonics. The flow harmonics, and especially the first and second harmonic (directed and elliptic flow) are known to be sensitive tools for the investigation of the equation-of-state.   
Generally, the flow coefficients $v_n$ arise from the Fourier series expansion of the angular part of the total particle spectra as
\begin{equation}
    E\frac{\mathrm{d}^3N}{\mathrm{d}p^3} = \frac{1}{2\pi} \frac{\mathrm{d}^2N}{p_\mathrm{T}\mathrm{d}p_\mathrm{T}\mathrm{d}y} \frac{\mathrm{d}N}{\mathrm{d}\phi}
\end{equation}
in which the angular part can be written as the infinite sum
\begin{equation}
    \frac{\mathrm{d}N}{\mathrm{d}\phi} = 1 + 2\sum\limits_{n=1}^\infty v_n\cos(n(\phi - \Psi_{\rm RP})),
\end{equation}
where $v_n$ is the n-th harmonic (Fourier) flow coefficient and $\Psi_{\rm RP}$ is the angle of the reaction plane. In line with the method employed by the HADES experiment which estimates the event plane from the spectator reaction plane, we use $\Psi_{\rm RP}=0$ in the simulation. The magnitude and sign of the extracted coefficients then give insights about the shape (in momentum space) of the expanding fireball in transverse direction. The coefficients are calculated as $v_n=\langle\cos(n\phi)\rangle$ in which the average $\langle\cdot\rangle$ is taken over all free protons, deuterons or pions in the events at a given rapidity and/or transverse momentum bin.

\begin{figure} [t!hb]
    \centering
    \includegraphics[width=\columnwidth]{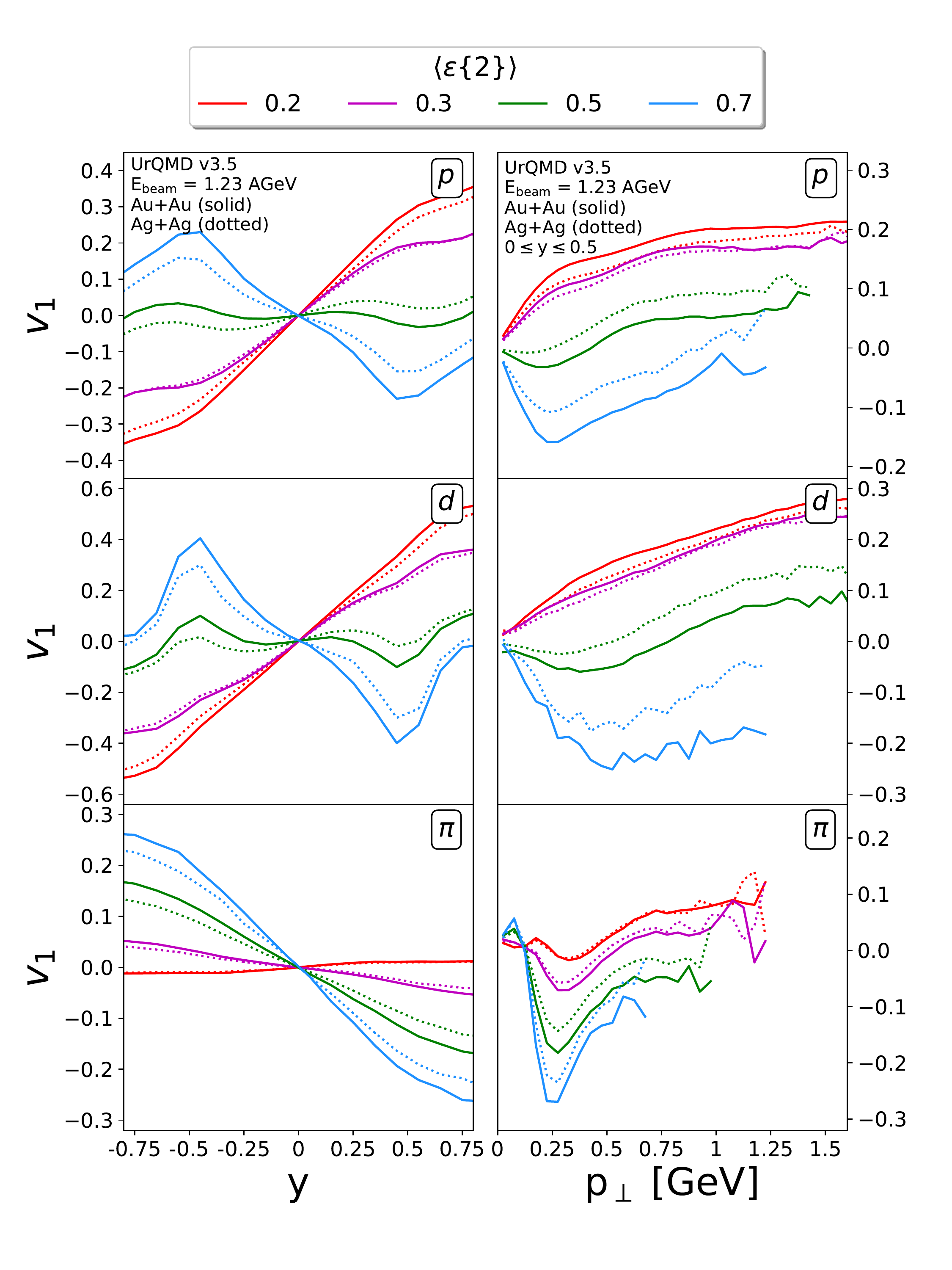}
    \caption{[Color online] The rapidity dependence (left column) and the transverse momentum dependence (right column) of the directed flow $v_1$ of protons (upper row), deuterons (middle row) and pions (lower row) from Ag+Ag reactions (dotted lines) and Au+Au reactions (solid lines) at $E_\mathrm{lab}$ = 1.23 $A$GeV and at fixed average transverse overlap eccentricities $\langle\varepsilon\{2\}\rangle$ = 0.2 (red), 0.3 (magenta), 0.5 (green) and 0.7 (blue) as defined above from UrQMD.}
    \label{fig:v1_y_pt}
\end{figure}

\begin{figure} [t!hb]
    \centering
    \includegraphics[width=\columnwidth]{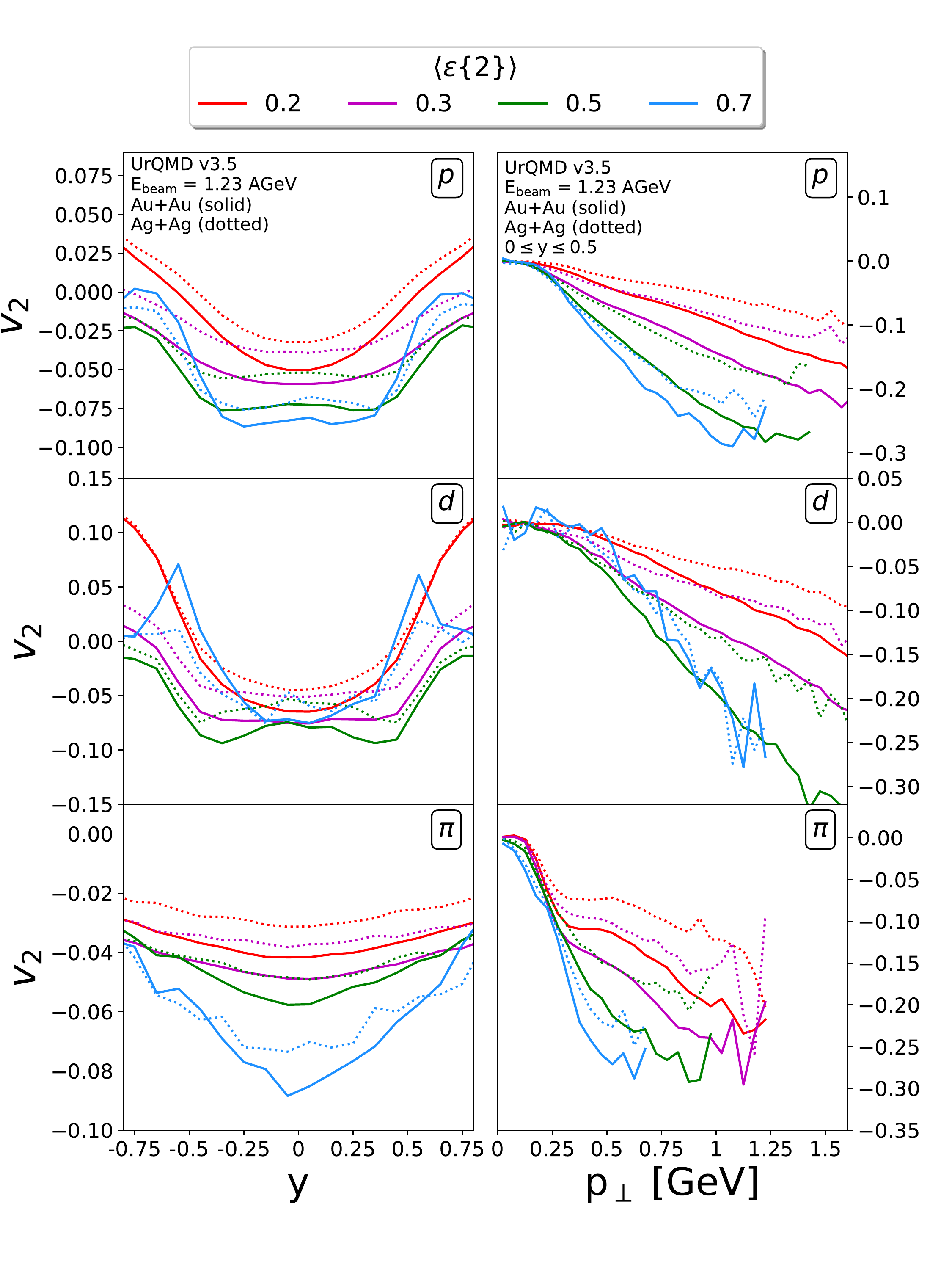}
    \caption{[Color online] The rapidity dependence (left column) and the transverse momentum dependence (right column) of the elliptic flow $v_2$ of protons (upper row), deuterons (middle row) and pions (lower row) from Ag+Ag reactions (dotted lines) and Au+Au reactions (solid lines) at $E_\mathrm{lab}$ = 1.23 $A$GeV and at fixed average transverse overlap eccentricities $\langle\varepsilon\{2\}\rangle$ = 0.2 (red), 0.3 (magenta), 0.5 (green) and 0.7 (blue) as defined above from UrQMD.}
    \label{fig:v2_y_pt}
\end{figure}
From very high energetic collisions it is well known that a scaling of the momentum space flow harmonic $v_n$ with the spatial eccentricity $\varepsilon_n$ can be expected, i.e. $v_n = \alpha \varepsilon_n$. The parameter $\alpha$ encodes the response of the system to the spatial anisotropy and is at high energies typically of the order $0.1-0.2$ \cite{PHOBOS:2006fqf}. In contrast, at the low energies investigated in this article, the elliptic flow of nucleons is negative at midrapidity because of significant spectator shadowing. Recently, the HADES collaboration has reported on a first observation of scaling of the elliptic flow with the initial transverse overlap eccentricity \cite{HADES:2022osk} with $\alpha=-0.2$ to $-0.4$. Although scaling is observed at very high energies and also at very low energies, the different signs of the scaling parameter strongly suggests a different physical reason. Thus, a meaningful comparison of the higher order flows between different collision systems should be done at the same initial transverse overlap eccentricity.

The eccentricity classes have already been introduced above and defined in Tab. \ref{tab:b-a-eps}. Here we compare the flows for mid-peripheral reactions, i.e. for eccentricities of 0.2, 0.3, 0.5 and 0.7, corresponding to impact parameters of $b=6.0, 8.2, 11.1, 13.1$ fm (for the Au+Au system) and $b=4.4, 6.1, 8.4, 10.3$ fm (for the Ag+Ag system). 

\subsection{Comparison of the directed flow $v_1$}
Let us start with a comparison of the directed flow $v_1$. In the left column of Fig. \ref{fig:v1_y_pt}, the rapidity dependence of the directed flow $v_1$ is shown for protons (top left), deuterons (middle left) and pions (bottom left) from Ag+Ag reactions (dotted lines) and Au+Au reactions (solid lines) at $E_\mathrm{lab}$ = 1.23 $A$GeV and at fixed average transverse overlap eccentricities $\langle\varepsilon\{2\}\rangle$ = 0.2 (red), 0.3 (magenta), 0.5 (green) and 0.7 (blue).

The first observation is that at fixed average transverse overlap eccentricity the qualitative shape of the directed flow $v_1$ extracted from Ag+Ag collisions is very similar to the one from Au+Au collisions at the same collision energy and for the investigated hadrons. This goes hand in hand with system-size independence of directed flow observed by STAR at larger center-of-mass energies \cite{STAR:2008jgm}. They are even comparable on a quantitative basis with deviations on the order of only 10\% between the two systems. However, the deviation becomes larger with increasing $\langle\varepsilon\{2\}\rangle$, i.e. towards more peripheral reactions. An interesting observation is that the flow of the nucleons and deuterons turns its sign for peripheral reactions, indicating that the remaining spectator matter does not receive a substantial push (which is usually responsible for the positive $v_1$ at forward rapidities), but acts only as an absorber leading to a net emission of the protons and deuterons towards the negative x-direction. Such shadowing is also present for pions for the three most central eccentricities, but turns to the opposite for the most peripheral eccentricity ($\langle\varepsilon\{2\}\rangle$ = 0.7). The reason is that the pions for the most peripheral reactions are not independently shadowed from the nucleons but follow the flow of the nucleon resonances, because they emerge from the nucleon resonance by their later decay.

The right column of Fig. \ref{fig:v1_y_pt} show the transverse momentum dependence of the directed flow $v_1$ of protons (top right), deuterons (middle right) and pions (bottom right) from Ag+Ag reactions (dotted lines) and Au+Au reactions (solid lines) at $E_\mathrm{lab}$ = 1.23 $A$GeV and at fixed average transverse overlap eccentricities $\langle\varepsilon\{2\}\rangle$ = 0.2 (red), 0.3 (magenta), 0.5 (green) and 0.7 (blue) as defined above from UrQMD.

Here, again, the qualitative as well as the quantitative shape and values of the transverse momentum dependence of the directed flow $v_1$ between the silver+silver and the gold+gold collision systems match to a high degree towards more central reactions. For peripheral reaction one observes again that influence of the shadowing of the spectators leads to a more negative $v_1$ at slightly forward rapidities ($0<y<0.5$).

\subsection{Elliptic flow $v_2$}
Finally, we analyze the elliptic flow $v_2$. In Fig. \ref{fig:v2_y_pt} (left column) we show the elliptic flow  $v_2$ as a function of rapidity for protons (top left), deuterons (middle left) and pions (bottom left) from Ag+Ag reactions (dotted lines) and Au+Au reactions (solid lines) at $E_\mathrm{lab}$ = 1.23 $A$GeV, again at fixed average transverse overlap eccentricities $\langle\varepsilon\{2\}\rangle$ = 0.2 (red), 0.3 (magenta), 0.5 (green) and 0.7 (blue) as defined above. Qualitatively, the elliptic flow distribution follows the expectations: At midrapidity, the $v_2$ is strongly negative and increases towards forward and backward rapidities. One further observes that for both systems, $v_2$ becomes more negative with increasing eccentricity. However, quantitatively a comparison between both systems is rather difficult. Generally, the flow in the smaller Ag+Ag system is weaker (i.e. less negative at midrapidity) than in Au+Au. This difference is mainly attributed to the substantially stronger shadowing by the spectators in the Au-system than in the Ag-system, at the same eccentricity.

This observation is supported by the right column of Fig. \ref{fig:v2_y_pt} which shows the transverse momentum dependence of the elliptic flow $v_2$ of protons (top right), deuterons (middle right) and pions (bottom right) for the same eccentricity selection as above. Except for very peripheral reactions ($\langle\varepsilon\{2\}\rangle$ = 0.7), the Au+Au reaction shows substantially stronger negative elliptic flow for all explored hadrons than the Ag+Ag system. Especially in the light of the weaker stopping of the Au+Au reaction discussed above it seems questionable, if this difference can be attributed to a higher pressure created in the Au+Au system. It seems more likely that the large remaining spectators provide substantial shadowing of any expansion in the x-direction, resulting in more negative $v_2$ at higher transverse momenta. 

\begin{figure} [t!hb]
    \centering
    \includegraphics[width=\columnwidth]{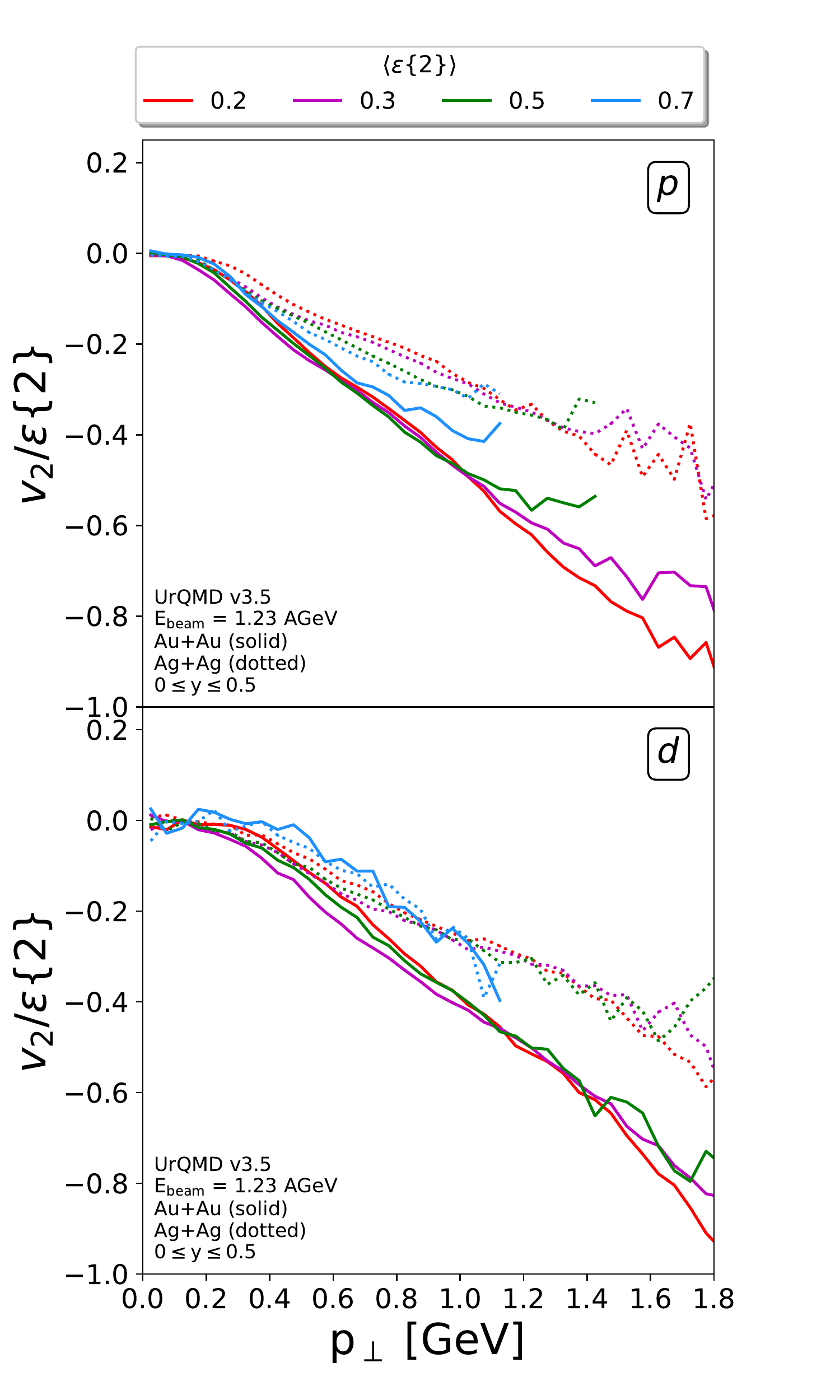}
    \caption{[Color online] The transverse momentum dependence of the elliptic flow $v_2$ divided by the transverse overlap eccentricity $\langle\varepsilon\{2\}\rangle$ of protons (upper row) and deuterons (lower row) from Ag+Ag reactions (dotted lines) and Au+Au reactions (solid lines) at $E_\mathrm{lab}$ = 1.23 $A$GeV and at fixed average transverse overlap eccentricities $\langle\varepsilon\{2\}\rangle$ = 0.2 (red), 0.3 (magenta), 0.5 (green) and 0.7 (blue) as defined above from UrQMD.}
    \label{fig:v2_eps_pt}
\end{figure}

To better understand the scaling with the transverse overlap eccentricity we next show in Fig. \ref{fig:v2_eps_pt} the transverse momentum dependence of the elliptic flow divided by the transverse overlap eccentricity $v_2/\langle\varepsilon\{2\}\rangle$ of protons (upper row) and deuterons (lower row) from Ag+Ag reactions (dotted lines) and Au+Au reactions (solid lines) at $E_\mathrm{lab}$ = 1.23 $A$GeV and at fixed average transverse overlap eccentricities $\langle\varepsilon\{2\}\rangle$ = 0.2 (red), 0.3 (magenta), 0.5 (green) and 0.7 (blue) as defined above from UrQMD. We observe in line with our argumentation that the transverse momentum dependence of the ratio $v_2/\langle\varepsilon\{2\}\rangle$ is independent of the eccentricity. However, the gold-gold collisions show a stronger (more negative) scaled elliptic flow than the silver-silver system for the protons and the deuterons.  

\begin{figure} [t!hb]
    \centering
    \includegraphics[width=\columnwidth]{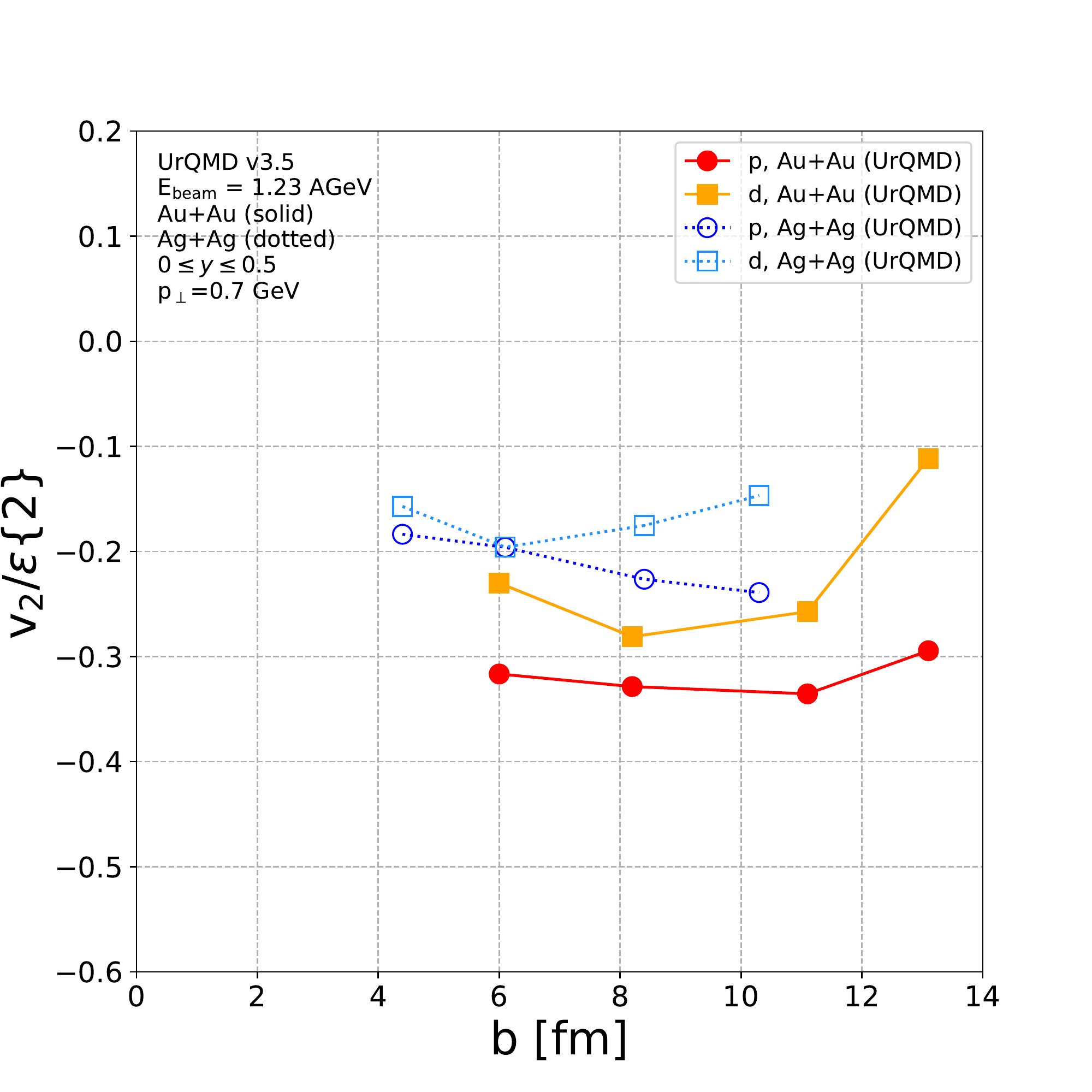}
    \caption{[Color online] The impact parameter (centrality) dependence of the elliptic flow $v_2$ divided by the transverse overlap eccentricity $\langle\varepsilon\{2\}\rangle$ of protons and deuterons from Ag+Ag reactions ($p$: open  blue circles, $d$: open light blue squares) and from Au+Au reactions ($p$: full red circles, $d$: full orange squares) taken at p$_\perp=0.7$ GeV and in $|y|\leq0.5$ at $E_\mathrm{lab}$ = 1.23 $A$GeV from UrQMD.}
    \label{fig:v2_eps_b}
\end{figure}

To bridge the gap to recently measured flow data \cite{HADES:2022osk} we lastly investigate the scaled elliptic flow as a function of impact parameter (centrality). For this Fig. \ref{fig:v2_eps_b} shows the impact parameter (centrality) dependence of the elliptic flow $v_2$ divided by the transverse overlap eccentricity $\langle\varepsilon\{2\}\rangle$ of protons and deuterons from Ag+Ag reactions ($p$: open  blue circles, $d$: open light blue squares) and from Au+Au reactions ($p$: full red circles, $d$: full orange squares) taken at p$_\perp=0.7$ GeV and in $|y|\leq0.5$ at $E_\mathrm{lab}$ = 1.23 $A$GeV from UrQMD. 

We observe two features: First, the impact parameter (centrality) dependence of the scaled elliptic flow for each particle species is more or less constant in the simulation except the most peripheral collisions in the Au+Au system which show a slight increase. Secondly, we observe that the silver-silver collisions (open symbols) show a smaller value of v$_2/\varepsilon\{2\}$ (i.e. closer to zero) than the gold-gold system at equal transverse overlap eccentricity. We conclude that indeed the transverse overlap eccentricity is reflected in the final state elliptic flow and is mostly centrality independent. However, it is not independent on the system size. This comes from the fact that although a similar spatial overlap shape is used for both systems, spectator shadowing is stronger in the Au+Au system because the spectator ``wall" is thicker allowing for more absorption/reflection and thus producing a more negative elliptic flow.  

\section{Conclusion}
We presented predictions for the comparison of the recently measured Ag+Ag and Au+Au runs at a beam energy of $E_\mathrm{lab}$ = 1.23~$A$GeV explored by the HADES experiment. We discussed sensible centrality selections for a meaningful comparison of both systems. To this aim, we suggest that both systems should be compared at the same number of participating nucleons in case of particle production. With the centrality selection, we found that central Ag+Ag reactions and mid-central Au+Au reactions show similar deuteron yields per participant and similar pion to proton ratios when properly scaled with $A_\mathrm{part}$. However, towards central Ag+Ag reactions, the scaling is violated. This scaling violation for very central Ag+Ag reactions in comparison to mid-peripheral Au+Au reaction (both at the same $A_\mathrm{part}$) was also found in the case of the rapidity densities and transverse momentum distributions. We related this difference to the weaker stopping power in mid-peripheral Au+Au reactions than in very central Ag+Ag collisions.

To explore the expansion dynamics in more detail, we investigated the directed and elliptic flow in terms of the flow harmonics. Here we suggested a centrality selection based on the initial state eccentricity to compare both systems. Such centrality selection is known to work well at high energetic collisions. We observed that for such selected geometries the elliptic flow of Au+Au and Ag+Ag reactions is qualitatively comparable with each other. However, a quantitative comparison seems difficult, due to the different shadowing strength in the x-direction in both collision systems which leads to a different scaling factor. 

\begin{acknowledgements}
The authors thank Paula Hillmann and Behruz Kardan for fruitful discussion about the flow harmonics and the analysis. This article is part of a project that has received funding from the European Union’s Horizon 2020 research and innovation program under grant agreement STRONG – 2020 - No 824093. J.S. thanks the Samson AG for funding. Computational resources were provided by the Center for Scientific Computing (CSC) of the Goethe University and the ``Green Cube" at GSI, Darmstadt. This research has received funding support from the NSRF via the Program Management Unit for Human Resources \& Institutional Development, Research and Innovation [grant number B16F640076].
\end{acknowledgements}



\end{document}